\documentclass[sigconf, screen, nonacm]{acmart}
\acmBooktitle{}
\acmPrice{}
\acmISBN{}
\acmDOI{}

\renewcommand\footnotetextcopyrightpermission[1]{} 

\usepackage{amsmath,amssymb}
\usepackage{algorithmic}
\usepackage{textcomp}
\usepackage{xcolor}
\usepackage{listings}
\usepackage{minted}
\usepackage{subcaption}
\usepackage{booktabs}
\usepackage{multirow}
\usepackage{makecell}
\usepackage{geometry}
\usepackage{array}
\usepackage{caption}
\usepackage{graphicx} 
\usepackage{float}
\usepackage{adjustbox}
\usepackage{marvosym}
\usepackage{ifsym}

\AtBeginDocument{%
  }

\begin{document}
\pagestyle{plain}

\title{Lyra: A Hardware-Accelerated RISC-V Verification Framework with Generative Model-Based Processor Fuzzing}

\author{
    Juncheng Huo$^{\ast1,2}$,
    Yunfan Gao$^{\ast1,2}$,
    Xinxin Liu$^{\ast3,4}$,
    Sa Wang$^{1,2}$,
    Yungang Bao$^{1,2}$,
    \\Xitong Gao$^{\dagger3,5}$ and
    Kan Shi$^{\dagger1,2}$\\[4pt]
    $^{1}$SKLP, Institute of Computing Technology, Chinese Academy of Sciences\\
    $^{2}$University of Chinese Academy of Sciences\\
    $^{3}$Shenzhen Institutes of Advanced Technology, Chinese Academy of Sciences\\
    $^{4}$Southern University of Science and Technology\\
    $^{5}$Shenzhen University of Advanced Technology\\[4pt]
}


\begin{abstract}
  As processor designs grow more complex, verification remains bottlenecked by slow software simulation and low-quality random test stimuli. Recent research has applied software fuzzers to hardware verification, but these rely on semantically blind random mutations that may generate shallow, low-quality stimuli unable to explore complex behaviors. These limitations result in slow coverage convergence and prohibitively high verification costs. In this paper, we present Lyra, a heterogeneous RISC-V verification framework that addresses both challenges by pairing hardware-accelerated verification with an ISA-aware generative model. Lyra executes the DUT and reference model concurrently on an FPGA SoC, enabling high-throughput differential checking and hardware-level coverage collection. Instead of creating verification stimuli randomly or through simple mutations, we train a domain-specialized generative model, LyraGen, with inherent semantic awareness to generate high-quality, semantically rich instruction sequences. Empirical results show Lyra achieves up to $1.27\times$ higher coverage and accelerates end-to-end verification by up to $107\times$ to $3343\times$ compared to state-of-the-art software fuzzers, while consistently demonstrating lower convergence difficulty.
\end{abstract}



\maketitle

\begingroup
\renewcommand\thefootnote{}
\footnotetext{%
$^{\ast}$These authors contributed equally.\\
$^{\dagger}$Corresponding authors. Emails: \texttt{xt.gao@siat.ac.cn}, \texttt{shikan@ict.ac.cn}
}
\endgroup

\section{Introduction}
As processor designs grow more complex, verification has become a critical part of the chip development cycle. Recent studies show verification consumes up to 70\% of development efforts \cite{arch_sim_survey}. The emergence of new Instruction Set Architectures (ISAs) like RISC-V has increased demand for more agile and efficient verification methods. This has driven growing interest in novel architectural modeling and efficient simulation methods, ensuring that open-source CPUs and IPs are reliable for community use.

\begin{figure}
\centering
\begin{subfigure}[b]{0.46\textwidth}
\includegraphics[width=\textwidth]{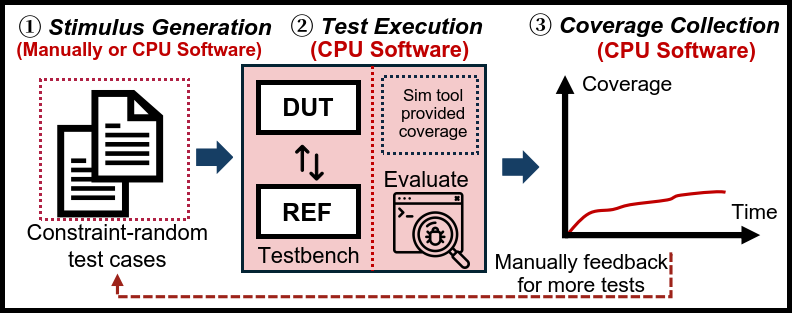}
\caption{Traditional verification flow.}
\includegraphics[width=\textwidth]{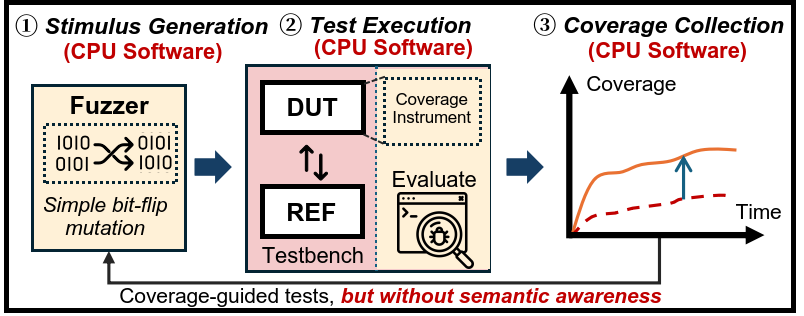}
\caption{Previous approach with software fuzzers.}
\includegraphics[width=\textwidth]{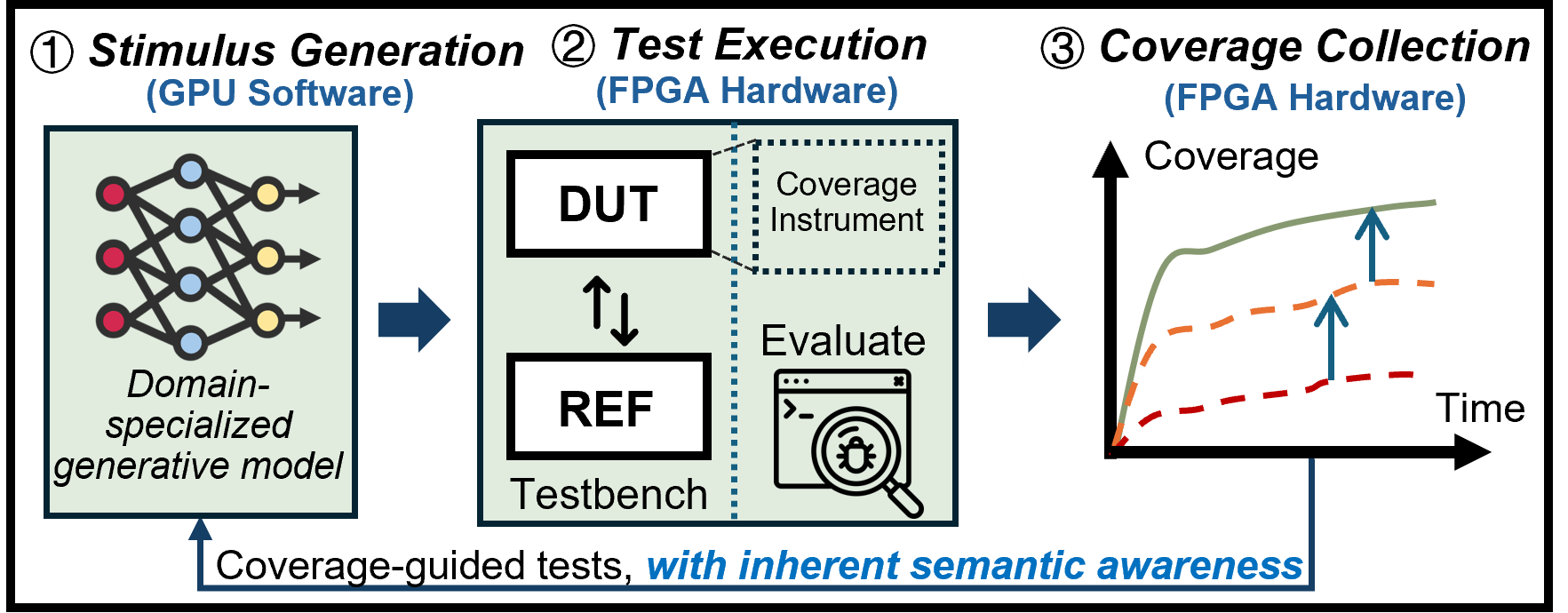}
\caption{Proposed verification flow using Lyra.}
\end{subfigure}
\caption{ (a) Traditional verification flow: all steps implemented in CPU software. (b) Advanced verification flow with software-based fuzzers: enables coverage-guided mutation for better coverage, but suffers from performance bottlenecks and semantic limitations. (c) Proposed {\em Lyra}: a heterogeneous verification framework combining a domain-specialized generative model with FPGA acceleration.}
\label{fig:first_page}
\end{figure}

However, conventional verification efficiency remains limited by two major challenges. First, conventional verification methodologies rely heavily on software simulation. The three main phases of the verification cycle: stimulus generation, test execution of both design-under-test (DUT) and its reference model (REF), and coverage collection are all performed in software, as shown in Figure~\ref{fig:first_page}~(a). While this provides fine-grained simulation visibility, the performance is extremely low, typically reaching only tens of kHz.

The second challenge is that verification effectiveness depends heavily on human experience, particularly in test case construction. Verification engineers must manually create numerous test cases with constrained-random (CR) stimulus, hoping to cover certain features. However, this approach is time-consuming and ineffective, as manually crafting specific combinations of complex instructions to trigger corner cases is extremely difficult.

To generate more intelligent test inputs, research has adopted software fuzzing techniques for processor verification~\cite{RFUZZ,Hyperfuzzing,ProcessorFuzz,Effective_processor_verification,TheHuzz,Cascade}, as shown in Figure~\ref{fig:first_page}~(b). Unlike traditional CR-based approaches, fuzzing uses coverage feedback to guide test generation. The fuzzer generates initial inputs, measures their coverage, and then mutates them to explore new states while avoiding redundancy.

However, existing fuzzers typically rely on blind, random mutations such as bit flips and lack a deep understanding of instruction semantics. While this approach may cover shallow verification scenarios quickly, hitting deeply buried corner cases requires logically precise and semantically coherent instruction sequences. Such meaningful sequences are exceedingly unlikely to emerge from purely random mutations, making this semantic blindness a fundamental bottleneck that limits the effectiveness of traditional fuzzers.

To address these two challenges, we propose Lyra, an efficient RISC-V verification framework with FPGA acceleration and generative model-based processor fuzzing, as shown in Figure~\ref{fig:first_page}~(c). Lyra implements both the DUT and the REF on the same FPGA SoC: the DUT runs on programmable logic while the REF runs on hardened ARM processors. Dedicated hardware checkers perform runtime differential checking of execution results at the instruction level. Lyra incorporates the control register-based coverage instrumentation method widely used in state-of-the-art fuzzing techniques~\cite{DifuzzRTL,Cascade} and implements it directly on FPGA. This enables the entire coverage collection to run on the FPGA itself.

To address the semantic limitations of software fuzzers, we train a domain-specialized generative model with 125 million parameters called LyraGen. This model generates high-quality RISC-V instruction streams while inherently understanding RISC-V semantics. We create novel encoding methods that translate instructions and their corresponding coverage information into tokens. Under such supervised learning, the model learns to generate instructions that achieve better coverage. During inference, we implement an instruction legality filter and an address checker to modify or remove illegally generated instructions, further improving verification effectiveness. In summary, the main contributions of this paper are as follows:

\begin{itemize}
    \item We design Lyra, the first heterogeneous GPU-CPU–FPGA co-verification framework. It offloads test execution, differential checking, and coverage collection to hardware while using a generative model for efficient stimulus generation.
    \item We develop LyraGen, a domain-specialized ISA-aware generative model, enabled by a novel RISC-V instruction tokenization scheme and supervised coverage-conditioned training, producing semantically rich instruction sequences that accelerate coverage convergence.
    \item We demonstrate empirically that Lyra achieves up to $1.27\times$ higher coverage, up to $107\times\sim3343\times$ faster end-to-end verification, and consistently lower convergence difficulty.
\end{itemize}

\section{Background and Related Works}

\subsection{Hardware Fuzzing}
Fuzzing originated in software testing as a technique that continuously generates and executes randomized test inputs to discover vulnerabilities in programs. Compared to traditional directed testing, fuzzing provides greater automation and effectively explores corner cases and rare execution paths that conventional methods struggle to reach. Inspired by its success in software, researchers have increasingly applied fuzzing to hardware verification. RFUZZ \cite{RFUZZ} was the first to adapt the American Fuzzy Lop (AFL) framework \cite{afl-fuzz} for hardware testing. DifuzzRTL \cite{DifuzzRTL} improved upon RFUZZ by adding clock-sensitive optimization and incorporating a reference model, enabling better capture of state transitions and more effective checking of RTL execution results. ProcessorFuzz \cite{ProcessorFuzz} introduced a coverage metric guided by processor Control and Status Register (CSR) transitions. Cascade \cite{Cascade} went further by employing a program-generation approach to construct longer test programs with complex control and data flows.

However, DifuzzRTL and ProcessorFuzz still suffer from low-quality test stimuli due to their mutation schemes and corpus scheduling approaches. Cascade avoids the overhead of output comparison by modifying its bug evaluation mechanism, but risks missing bugs as a trade-off. Moreover, all these frameworks lack awareness of instruction semantics and rely on software-level implementation, resulting in inefficiencies when verifying large and complex hardware designs such as CPU cores. In comparison, Lyra retrains a domain-specialized generative model for more efficient instruction generation while offloading time-consuming test execution and coverage collection entirely to an FPGA, significantly improving overall verification performance.

\subsection{ML-based Fuzzing}



ChatFuzz \cite{chatfuzz} is an ML–based hardware fuzzing framework that uses a GPT-2 model to interpret processor language and generate assembly code sequences, but depends on closed-source models and suffers from reproducibility issues due to LLM randomness. GenHuzz \cite{genhuzz} uses hardware-guided reinforcement learning with coverage feedback, but works at the assembly level and requires assembler translation, limiting instruction diversity and increasing computational overhead. Both works rely on VCS-based coverage metrics, creating a software simulation performance bottleneck.

Lyra addresses these issues through machine-code–oriented encoding and legality checking, enabling direct generation of diverse instructions while avoiding syntax learning difficulties. Coverage-conditioned training accelerates convergence and reduces local optima. Furthermore, Lyra uses FPGA-based execution acceleration, significantly improving end-to-end system performance.


\section{The Lyra Framework Overview}

\begin{figure*}[tb]
  \centering
  \includegraphics[width=.8\textwidth]{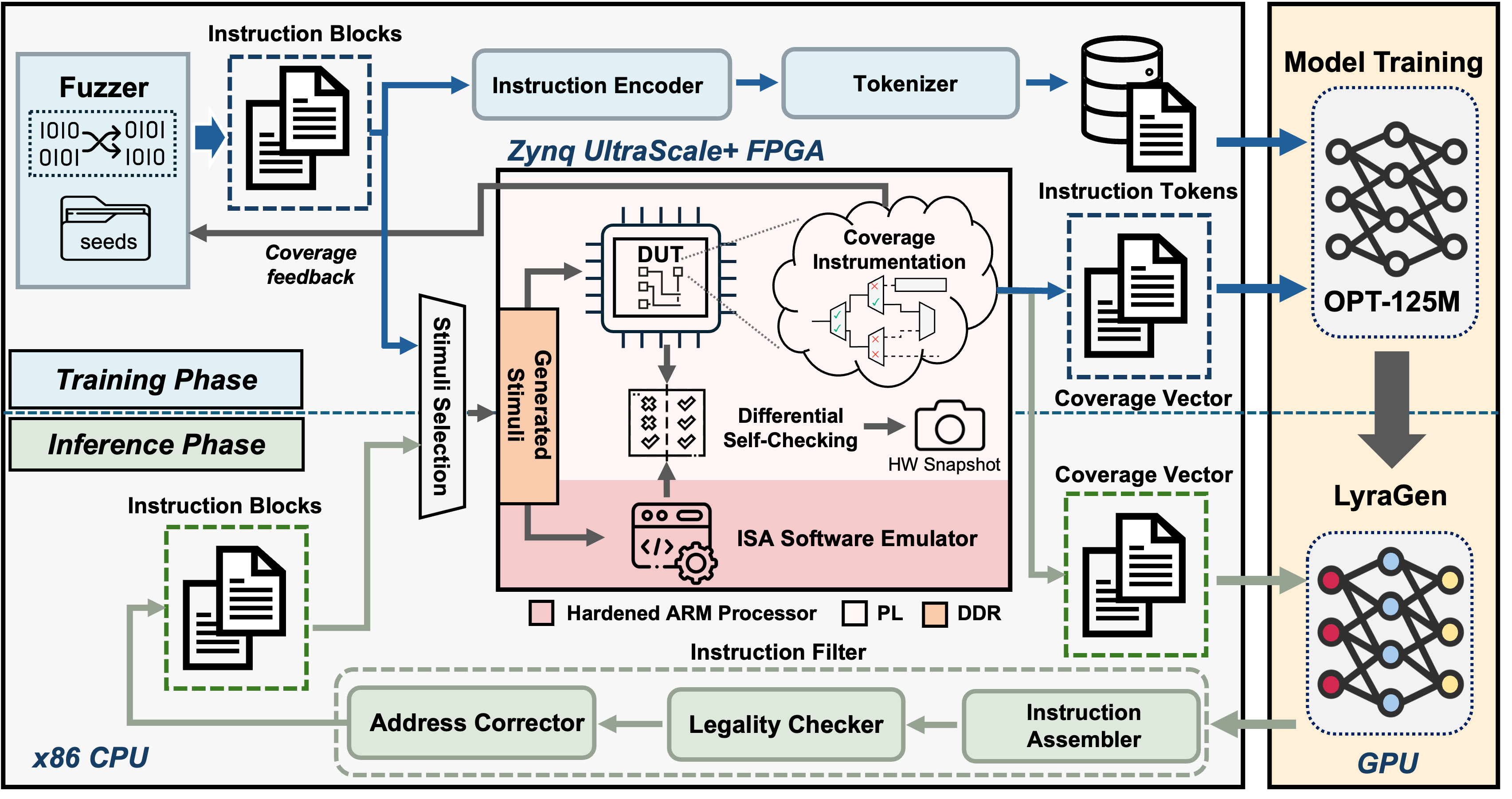}
  \caption{Overview of the Lyra framework, including the training phase and the inference phase, both running on the same heterogeneous system containing GPU, CPU and FPGA.}
  \label{Fig:overview}
\end{figure*}

Lyra is a heterogeneous verification framework that combines a domain-specialized generative model, LyraGen, for high-quality RISC-V instruction generation, which is deployed on a GPU; an FPGA-based testbed, and CPU-based software programs for post-processing model-generated instructions. The construction and usage of Lyra contains two phases, as shown in Figure~\ref{Fig:overview}.

\textbf{The training phase} aims to train LyraGen. Model training runs on the GPU, while the training datasets are generated via software-hardware co-simulation on the CPU and FPGA. Section~\ref{sec:training} describes the training phase in detail. Once trained, the model serves as the instruction-generation engine to drive RISC-V processor verification during \textbf{the inference phase}, detailed in Section~\ref{sec:inference}. Both phases share the same FPGA testbed rather than relying on time-consuming software simulation. The testbed includes a differential checking module based on the \textsc{Encore} structure~\cite{encore_fpga23}, which performs self-checking between the DUT and its reference model running concurrently on the same FPGA SoC.

Lyra automatically instruments synthesizable coverage points based on the Register Coverage metric—a widely used approach in recent hardware fuzzing works~\cite{DifuzzRTL}. This enables coverage collection to be performed directly on the FPGA.

\section{The Training Phase}\label{sec:training}
The training phase aims to develop a domain-specialized generative model, LyraGen, that generates high-quality RISC-V instruction streams with inherent semantic awareness. This overcomes the randomness and semantic blindness of traditional fuzzers. We redesign the encoding of RISC-V instructions so the model can effectively exploit their semantic structure. Building on this encoding scheme, we introduce a data-collection methodology and construct a training dataset containing numerous \texttt{<instruction, coverage>} entries.

We use OPT-125M~\cite{zhang2022optopenpretrainedtransformer} as the base model and extend it with architectural modifications and retraining to support the proposed encoding format and enable effective exploration of the coverage space. Through this process, LyraGen learns to generate large volumes of high-quality RISC-V instructions that align with the underlying ISA semantics.

\subsection{Instruction Encoding}
A non-compressed RISC-V instruction is 32 bits long and consists of multiple fields such as opcode, source and destination registers, and immediates that vary by instruction format. To help the model capture explicit semantic information and accelerate training convergence, we redesign the instruction representation by re-encoding each RISC-V instruction into a sequence of tokens as listed in Table~\ref{tab:field}. Each token is an integer with a maximum width of 8 bits, representing values up to 255.

Under this encoding scheme, an instruction is represented as a sequence of words, each of which serves as a token. The model learns the structural patterns of these token sequences and generates new tokens that can be reassembled into valid instructions. The tokenized content of each field follows the specification in Table~\ref{tab:field}, depending on the instruction type. The opcode field appears in all instructions, while other fields appear only in specific instruction categories. When a field does not exist for a given instruction, its token is omitted. Since each instruction format uniquely determines its required token set, these omissions introduce no ambiguity and the encoding forms a prefix code.

\begin{table}
  \centering
  \caption{Proposed Instruction Encoding and Tokenization}
  \label{tab:field}
  \resizebox{\columnwidth}{!}{%
    \begin{tabular}{ccccccc}
      \toprule
      \multirow{2}{*}{Type} & \multicolumn{6}{c}{Token ID \& Bit-width} \\
       & Token1 & Token2 & Token3 & Token4 & Token5 & Token6\\
      \midrule
      R & Opcode (7) & Funct7 (7) & Funct3 (3) & Rd (5) & Rs1 (5) & Rs2 (5) \\
      I & Opcode (7) & Funct3 (3) & Rd (5) & Rs1 (5) & ImmLo (8) & ImmMi (4) \\
      I (Shift) & Opcode (7) & Funct3 (3) & Rd (5) & Rs1 (5) & Shamt (7) & \\
      S & Opcode (7) & Funct3 (3) & Rs1 (5) & Rs2 (5) & ImmLo (8) & ImmMi (4) \\
      B & Opcode (7) & Funct3 (3) & Rs1 (5) & Rs2 (5) & ImmLo (8) & ImmMi (4) \\
      U & Opcode (7) & Rd (5) & ImmLo (8) & ImmMi (8) & ImmHi (4) & \\
      J & Opcode (7) & Rd (5) & ImmLo (8) & ImmMi (8) & ImmHi (4) & \\
      \bottomrule
    \end{tabular}
  }
\end{table}

Using this encoding scheme, we can systematically construct the model, build the dataset, train the network, and generate new instructions through inference.

\subsection{Generation of Training Data}
The objective of model training is to enable the model to learn the relationship between input verification vectors, i.e. the RISC-V instructions, and their corresponding coverage. By using high-quality instructions as ground truth, the model learns instruction characteristics within the input coverage vector space. This allows the model to generate high-quality verification vectors conditioned on a given coverage state. This approach improves overall coverage while avoiding the slow iteration speed and convergence difficulties typical of reinforcement-learning-based methods.

To generate high-quality \texttt{<instruction, coverage>} pairs, we construct a heterogeneous fuzzing system consisting of a software fuzzer running on a CPU and a hardware testbed implemented on an FPGA, as shown in the Training Phase of Figure~\ref{Fig:overview}. The software fuzzer connects to the DUT's main memory and continuously produces RISC-V instructions. These instructions are sent to the FPGA that integrates a differential self-checking mechanism. Following the \textsc{Encore} architecture, both the DUT and the software ISA emulator are instantiated on the same FPGA. Once the software fuzzer generates an instruction block and writes it into main memory, the DUT and ISA emulator simultaneously fetch and execute it.

To convert the instructions from the fuzzing system into the tokenized format our model requires, we add an encoding module at the output of the hardware fuzzing pipeline. This module translates each RISC-V instruction into the instruction-token representation.

We also instrument register-level coverage collection directly on the FPGA, enabling the system to extract coverage information immediately after each instruction executes and feed it back to the fuzzer. Guided by this feedback, the fuzzer mutates instructions to generate new candidates. For each executed instruction, its tokenized encoding and corresponding coverage vector are stored as a pair in the dataset for model training.

This supervised learning approach ensures the model acquires semantic knowledge of the RISC-V ISA and learns instruction contexts across diverse coverage scenarios, from low to high coverage. As a result, the model develops a comprehensive global view of coverage-space behavior while avoiding the inefficiency and instability of reinforcement learning.


\subsection{Training and Optimization of LyraGen}
To ensure fast computation and portability, we adopt OPT-125M as the base model and retrain it on the dataset described above. The model takes the 22-dimensional vector coverage data as input, and predicts the tokens of the next instruction. We modify the input modality of the original architecture: instead of receiving token IDs as conventional LLMs do, the model directly accepts a numerical vector. This vector is first processed through a linear layer rather than an embedding layer, and then fed into the model's hidden layers to produce the embedding representation for downstream computation. This modification allows the model to leverage the numerical semantics in the coverage vector, rather than treating these values as digit characters.

To accommodate diverse instruction token formats, we design an RVTokenizer whose vocabulary aligns with RISC-V instruction tokenization. The vocabulary contains 257 transformer tokens: one for each numerical value from 0 to 255, plus a padding symbol. Higher-bitwidth values, such as a 20-bit immediate, are represented as sequences of multiple tokens. This scheme avoids vocabulary explosion while preserving strong representational capacity.

Because the model's input interface and tokenizer were substantially redesigned and the resulting model no longer operates on natural language, we do not fine-tune the pretrained model. Instead, we train the modified model from scratch. Using the dataset described above, we train the model for 10 epochs on an NVIDIA GeForce RTX 4090 GPU, which takes 58 minutes total.

\section{The Inference Phase}\label{sec:inference}
The model outputs a sequence of tokens that must be reassembled into RISC-V instructions. However, due to the stochastic nature of probabilistic generative models, the token sequences do not always map to well-formed instructions. To address this, we design an instruction filter that verifies the syntactic and semantic validity of each reconstructed instruction during assembly.

For invalid instructions, we first attempt repair by mapping the token sequence to the closest valid instruction based on binary encoding similarity. If repair is not feasible, the instruction is discarded. For memory-access instructions, generated addresses may be out of bounds or misaligned, triggering exceptions that prevent subsequent instructions from executing. To mitigate this, the instruction filter includes an address-sanitization module that corrects faulty memory addresses.

After correction, instructions are fed into the FPGA testbed, which performs self-checking between the DUT and reference model. Hardware-synthesizable coverage points are automatically instrumented at compile time so coverage data can be collected directly on the FPGA at runtime and fed back into LyraGen to generate further instructions for coverage convergence.

\subsection{Instruction Legality Checker}
All generated tokens are sequentially assembled into initial instructions following the encoding scheme in Table~\ref{tab:field}. Each instruction then passes through a filter that checks whether it conforms to the RISC-V specification and will not trigger an illegal-instruction exception. If any issue is detected, the filter attempts to repair it.

The filter first reads the first two tokens to infer the instruction's base format (e.g., R-Type, I-Type). It then determines the opcode from the first token and generates the corresponding funct3 or funct7 fields. For the remaining fields, the filter applies different strategies based on their semantic constraints. If a field is restricted to a small set of legal values, the filter maps the generated token to one of these valid options via hash-based mapping. If a field may take any value within its bitwidth such as an immediate-field fragment, the filter uses the generated token value directly.

For immediates, our token representation splits the value into high, middle, and low segments to avoid vocabulary explosion. During legality checking, the filter reconstructs the full binary immediate by combining these segments according to the bitwidth required by the instruction type.

After all fields are validated and corrected, the filter assembles them into a complete RISC-V instruction in accordance with the ISA's standard encoding format.

\subsection{Instruction Address Correction}
Memory-access instructions require additional checking beyond the semantic legality check to prevent invalid or exceptional memory accesses. To address this, we design a memory-access address correction module based on a RISC-V ISA emulator. After model-generated instructions pass legality checking, they are simulated in the ISA Emulator. During simulation, the module inspects the target address whenever a memory-access instruction executes. If the address is invalid, it is corrected to fall within a predefined legal memory region, and the modified instruction is written back to memory. The emulator then refetches from the faulting instruction, allowing execution to resume normally. Since the ISA-level emulator does not model processor microarchitectural details, this process is significantly faster than hardware-level simulation.

Before applying corrections, the original instruction's fields, such as source registers, destination registers, and opcode, are preserved to ensure the address-fixing process does not alter the instruction's semantics. The correction strategy works in two steps: first, determine a legal target address; second, adjust the relevant instruction fields based on the difference between the original and corrected addresses. Two types of memory-access exceptions are handled: misalignment and out-of-bounds access. Misalignment occurs when the target address is not an integer multiple of the accessed data size. Out-of-bounds access occurs when the address falls outside the allocated memory region.

For misaligned accesses, the correction is straightforward: the lower bits of the target address are aligned to match the number of bytes accessed. For example, an \texttt{lh} instruction that loads two bytes requires the least significant bit of the address to be zero. The instruction's offset field is then adjusted to reflect the address change and maintain consistency.

Handling out-of-bounds accesses is more complex. The offset field is only 12 bits wide, so it cannot represent arbitrary 32-bit address adjustments. The difference between an illegal target address and the nearest legal address may exceed what the offset alone can correct. To address this, we introduce an auxiliary instruction sequence that adjusts the high bits of the address. The memory-access instruction and all subsequent instructions shift downward, and we fill the vacant slot with an \texttt{auipc} instruction followed by an \texttt{addi} instruction. Both instructions use the source and destination registers preserved from the original instruction. Their immediate values encode the difference between the high bits of the original and corrected addresses. After rewriting, execution restarts from the newly inserted \texttt{auipc}.


The instruction-legality checking ensures the validation process does not halt due to undefined or illegal instructions. The address-correction module prevents the processor core from accessing invalid memory regions. Together, these components significantly improve the effectiveness of generated instruction streams. 


\subsection{Fine-grained Self-checking with FPGA Acceleration}
Lyra eliminates the performance bottleneck of previous works that rely on time-consuming software simulations to collect coverage data from simulators such as VCS. Instead, post-processed instructions are fed into a differential checking module based on the \textsc{Encore} structure~\cite{encore_fpga23}, which runs on a modern FPGA System-on-Chip that combines programmable logic (PL) with hardened ARM processors. The DUT runs on the PL, while its software reference model runs on the on-chip ARM processors of the same FPGA. This setup enables concurrent execution with dynamic self-checking of I/O behaviors and key registers on both sides.

Lyra instruments synthesizable coverage points based on the Register Coverage metric~\cite{DifuzzRTL}. Coverage data is collected directly on the FPGA and fed back to LyraGen to guide further instruction generation, creating a fully automated verification loop with minimal human intervention.

Lyra also leverages hardware snapshot capabilities from the \textsc{Encore} structure to capture the complete FPGA state, including logic blocks, flip-flops, on-chip memories, and external devices such as DDR. Users can transfer these snapshots from the FPGA to analyze and debug issues using external software simulators like ModelSim on a host PC~\cite{attia2020statemover,sm_fpt}.

\section{Evaluation}

\subsection{Experimental Settings}
We implement Lyra on a workstation with an Intel Core i7-12700 CPU, NVIDIA GeForce RTX 4090 GPU, and Fidus Sidewinder FPGA board with an AMD Zynq UltraScale+ XCZU19EG FPGA. For the FPGA hardware implementation, we use Vivado 2020.2 with a 100MHz frequency target. We compare Lyra against two state-of-the-art software fuzzers DifuzzRTL and Cascade for RISC-V processor verification. All experiments use the open-source RISC-V processor RocketCore~\cite{rocket} as the DUT, with a total generation of 9.24 million \texttt{<instruction, coverage>} pairs for model training.

\subsection{Coverage Convergence Performance}
\begin{figure}[tb]
  \centering
  \includegraphics[width=.48\textwidth]{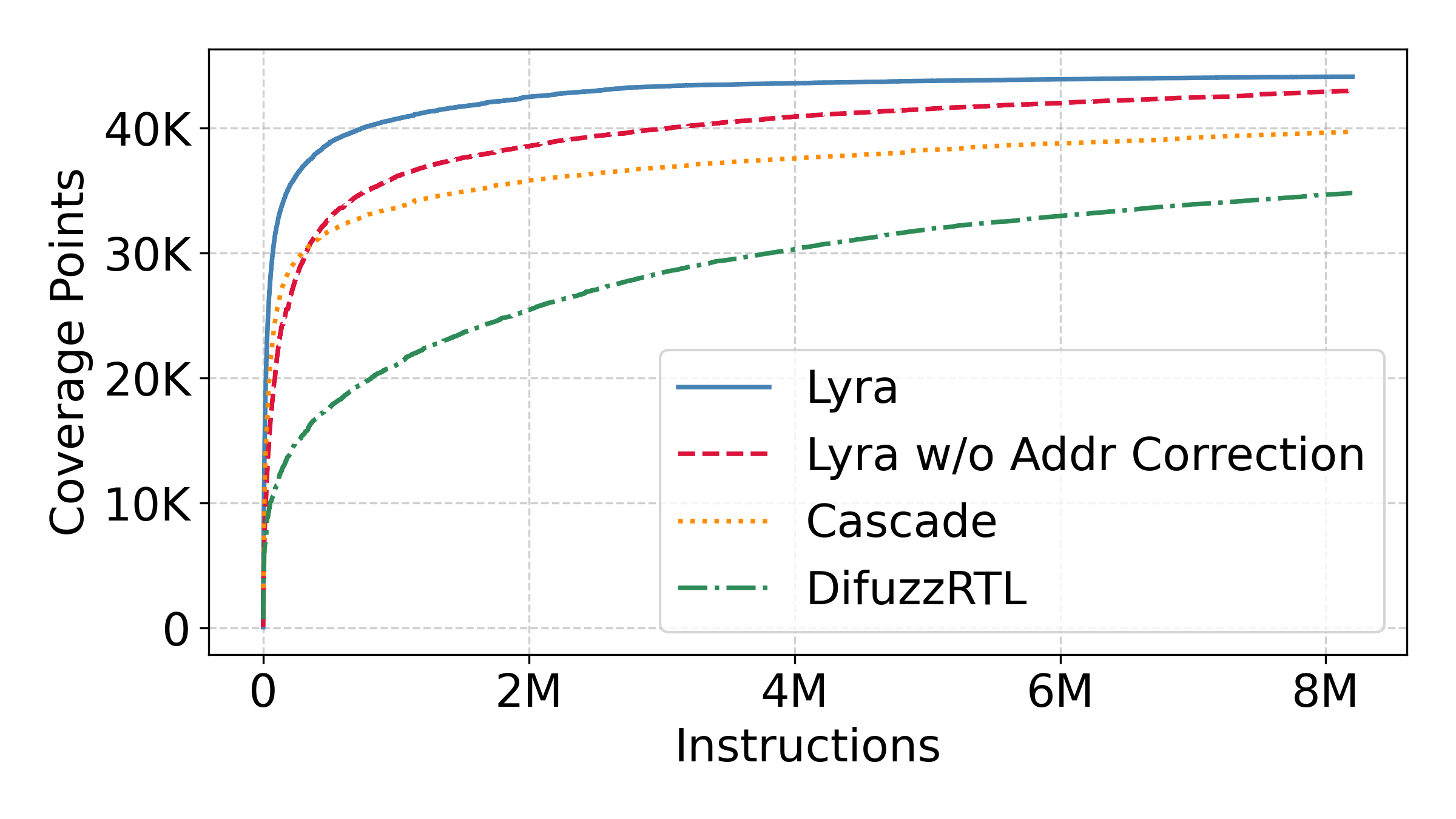}
  \caption{Comparison of coverage convergence performance between different methods.}
  \label{Fig:cov_inst_comp}
\end{figure}

We first evaluate the impact of ISA-emulator-based address correction by comparing coverage results with and without correction enabled. The results are shown in Figure~\ref{Fig:cov_inst_comp}.

It can be seen that the validation system with address correction (blue solid line) consistently achieves higher coverage than the system without it (red dotted line), given the same number of generated instructions. For example, at 10800 generated instructions, address correction yields $1.96\times$ higher coverage, which is the largest observed difference. Even as coverage begins to converge at 8 million instructions, the system with address correction still achieves $1.03\times$ higher coverage. Based on these results, we enable address correction by default in all subsequent experiments to maximize coverage.

We then compare the coverage convergence performance of Lyra, Cascade and DifuzzRTL, as shown in Figure~\ref{Fig:cov_inst_comp}. When executing the same number of instructions, Lyra achieves significantly higher coverage than both DifuzzRTL and Cascade. For example, after 1 million instructions, Lyra reaches 40738 coverage points—$1.21\times$ higher than Cascade (33610) and $1.94\times$ higher than DifuzzRTL (21030). At 8 million instructions, where coverage begins to converge, Lyra still outperforms the previous approaches by $1.11\times$ and $1.27\times$ against Cascade and DifuzzRTL, respectively. These results demonstrate that Lyra covers more states with fewer instructions and can potentially converge at a higher coverage level.

\subsection{End-to-end Performance}
\begin{figure}[tb]
  \centering
  \includegraphics[width=.48\textwidth]{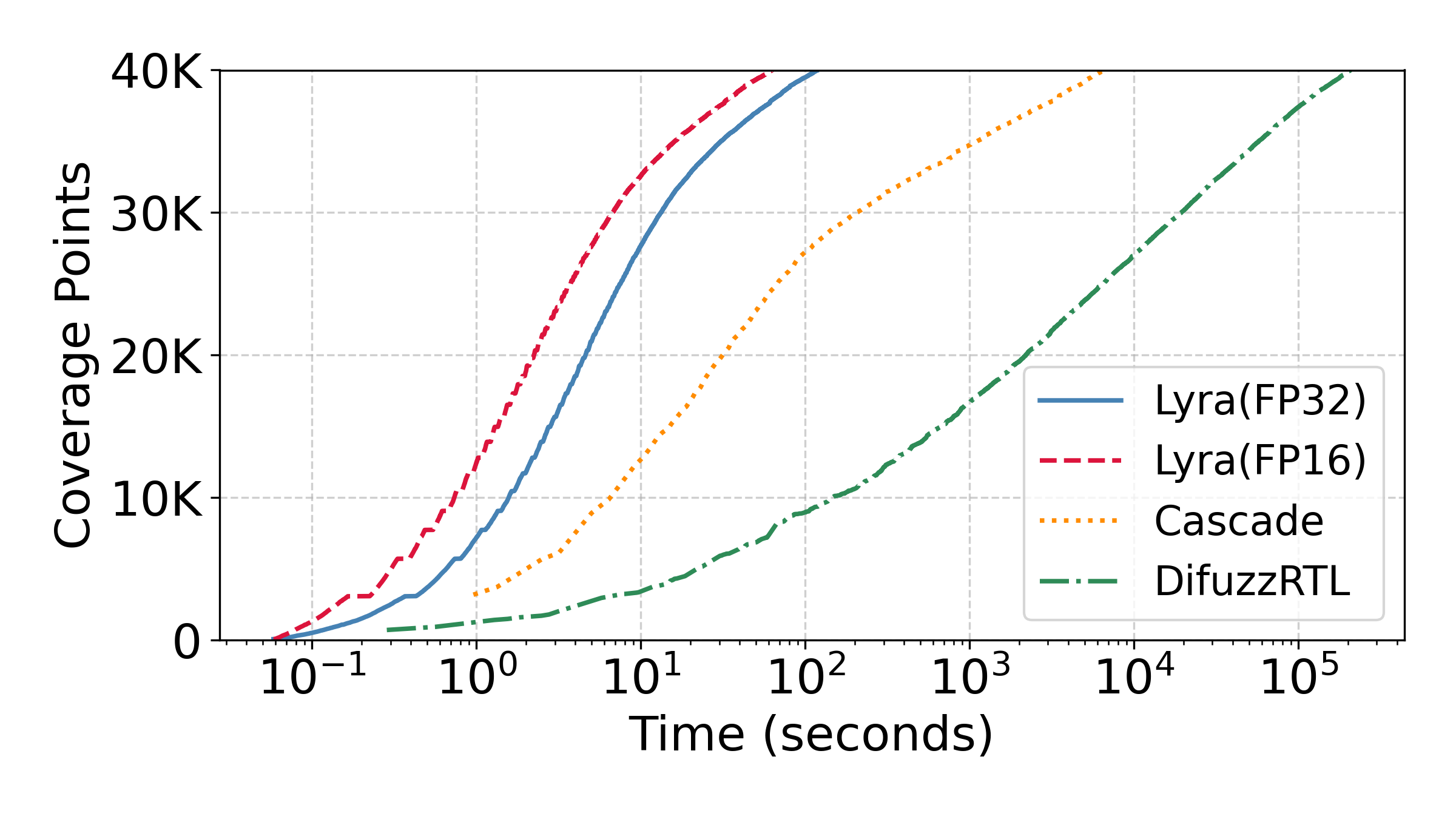}
  \caption{Comparison of different fuzzing methods to achieve certain coverage targets.}
  \label{Fig:cov_time_comp}
\end{figure}

For all three methods, we measure the end-to-end performance in time of the entire verification system, which includes instruction generation, post-processing if any (as in Lyra), and DUT execution. In addition to the full-precision version, we also deploy an FP16 implementation, as shown in Figure~\ref{Fig:cov_time_comp}. For a given coverage target, the proposed Lyra framework achieves significant performance improvements over previous approaches. For example, Lyra reaches 40000 coverage points in just 115.2 seconds, indicating a $57.4\times$ speed-up compared to Cascade’s 6610.9 seconds, and a $1797.3\times$ speed-up compared to DifuzzRTL's 207048.2 seconds. In addition, Lyra-FP16 achieves $1.86\times$ performance improvement over the full-prevision implementation, corresponding to a maximum $106.8\times$ and $3342.9\times$ speedup over Cascade and DifuzzRTL, respectively.

We profile the time consumption of each stage in the end-to-end workflow, instruction generation, post-processing, and instruction execution, and compute the overall system throughput, as summarized in Table~\ref{tab:perf_breakdown}. The performance improvement of Lyra stems from two key sources. First, LyraGen rapidly generates large volumes of high-quality instructions, enabling coverage to rise quickly to the target level. Second, FPGA-based hardware simulation greatly accelerates the execution. With full-precision inference, Lyra achieves system throughput $4.65\times$ and $48.1\times$ higher than Cascade and DifuzzRTL, respectively. With FP16 inference, Lyra's throughput increases by $1.86\times$ over its full-precision configuration, reaching $8.65\times$ and $89.5\times$ the throughput of Cascade and DifuzzRTL.

\begin{figure}[tb]
  \centering
  \includegraphics[width=.46\textwidth]{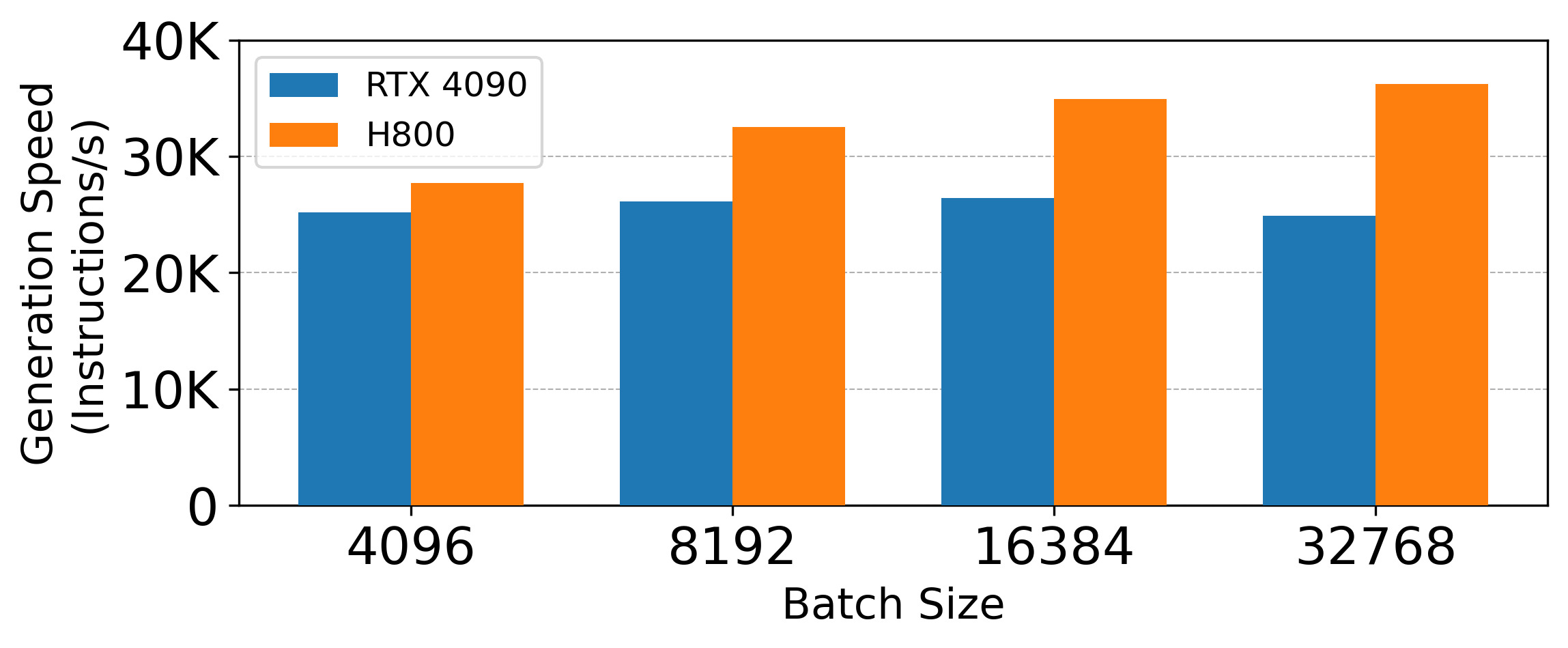}
  \caption{Comparison of LyraGen generation performance with different GPU hardware.}
  \label{Fig:h800}
\end{figure}

The design and implementation of LyraGen fully exploits the parallelism offered by modern GPUs. This means the model's performance can be further improved when deployed on more powerful GPU hardware. In contrast, CPU-based software fuzzers are fundamentally constrained by single-core CPU performance, making it difficult to substantially accelerate instruction generation through multicore parallelism. We deploy LyraGen-FP16 on an NVIDIA H800 GPU and evaluate its generation throughput under different batch sizes, as shown in Figure~\ref{Fig:h800}. Compared with the RTX 4090, generation performance improves by $1.1\times$ to $1.4\times$, demonstrating Lyra's strong scalability on higher-end GPU platforms.

\begin{table}[t]
\centering
\caption{Performance Breakdown of Different Methods}
\label{tab:perf_breakdown}
\adjustbox{max width=\linewidth}{
\begin{tabular}{@{}lcccc@{}}
\toprule
\textbf{Breakdown} &
\textbf{DifuzzRTL} &
\textbf{Cascade} &
\textbf{Lyra\_FP32} &
\textbf{Lyra\_FP16} \\
\midrule
Generation   & 16711.6s (CPU) & 522.8s (CPU) & 81.5s (GPU) & \textbf{\boldmath 28.2s (GPU)} \\
\midrule
Post-process   & N/A & N/A & 14.4s (CPU) &  14.4s (CPU) \\
\midrule
Execution    & 190271.6s (CPU) & 6058.1s (CPU) & \textbf{\boldmath 19.3s (FPGA)} &  \textbf{\boldmath 19.3s (FPGA)} \\
\midrule
Total        & 207048.2s & 6610.9s & 115.2s &  \textbf{\boldmath 61.9s} \\    
\midrule
\midrule
Throughput & 133.9 & 1385.4 & 6442.8 & \textbf{\boldmath 11990.5} \\
(Instructions/sec)\hspace{-2em} & $1\times$ & $10.3\times$ & $48.1\times$ & \textbf{\boldmath $89.5\times$} \\
\bottomrule
\end{tabular}}
\end{table}

\subsection{Analysis of Coverage Convergence Difficulty}
As coverage increases, further improvements become increasingly difficult. This challenge is especially pronounced in processor verification, where reaching certain deep microarchitectural states requires highly structured and semantically precise instruction sequences. Traditional fuzzers rely largely on blind, random mutations, making it extremely difficult to capture or exploit the complex semantics of instructions. In contrast, generative model-based approaches such as LyraGen intelligently generate high-quality validation vectors, accelerating coverage convergence.

\begin{figure}[tb]
  \centering
  \includegraphics[width=.48\textwidth]{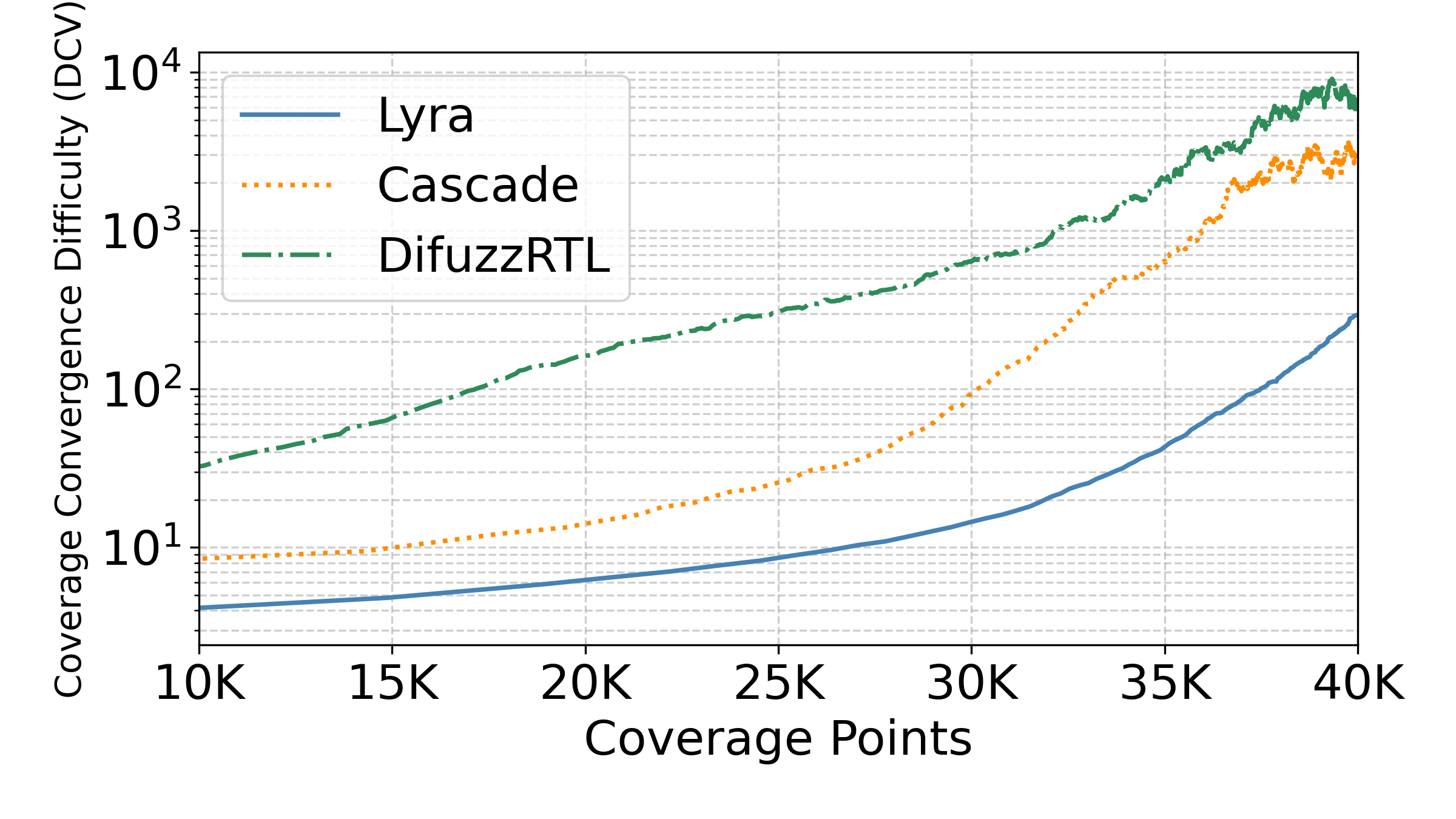}
  \caption{Comparison of coverage convergence difficulty between different methods.}
  \label{Fig:converge_diff}
\end{figure}

To quantify the difficulty of achieving coverage convergence, we define a metric $DCV=\Delta {Inst}/\Delta {Cov}$, where $\Delta {Inst}$ denotes the increase in executed instructions and $\Delta {Cov}$ represents the corresponding growth in coverage. A higher DCV value indicates that more instructions are needed to achieve each unit of coverage gain, reflecting greater difficulty in improving coverage. Experimental results in Figure~\ref{Fig:converge_diff} show that all methods face increasing difficulty as coverage rises. However, at any given coverage level, our method consistently achieves a much lower DCV than both DifuzzRTL and Cascade. This advantage becomes more pronounced at higher coverage levels. When coverage reaches 40K, the DCV values for Lyra, Cascade, and DifuzzRTL are 291.5, 2947.0, and 5607.6, respectively. This gap shows that Lyra faces lower convergence difficulty than baseline methods, enabling it to maintain faster convergence rates.


\section{Conclusion}
We present Lyra, a heterogeneous GPU-CPU-FPGA verification framework that addresses the semantic blindness of existing fuzzers and the performance limitations of software-based simulation. By combining ISA-aware instruction generation with hardware accelerated execution and checking, Lyra achieves up to $1.96\times$ higher coverage and upto $3343\times$ faster throughput than SOTA approaches.


\bibliographystyle{ACM-Reference-Format}
\bibliography{sample-base}

@inproceedings{encore_fpga23,
author = {Shi, Kan and Xu, Shuoxiang and Diao, Yuhan and Boland, David and Bao, Yungang},
title = {{ENCORE}: Efficient Architecture Verification Framework with {FPGA} Acceleration},
year = {2023},
isbn = {9781450394178},
url = {https://doi.org/10.1145/3543622.3573187},
doi = {10.1145/3543622.3573187},
booktitle = {Proceedings of the ACM/SIGDA International Symposium on Field Programmable Gate Arrays (FPGA)},
pages = {209–219},
numpages = {11},
location = {Monterey, CA, USA}
}

@article{arch_sim_survey,
author = {Akram, Ayaz and Sawalha, Lina},
year = {2019},
month = {05},
pages = {1-1},
title = {A {S}urvey of {C}omputer {A}rchitecture {S}imulation {T}echniques and {T}ools},
volume = {PP},
journal = {IEEE Access},
doi = {10.1109/ACCESS.2019.2917698}
}

@inproceedings{attia2020statemover,
  title={State{M}over: {C}ombining simulation and hardware execution for efficient {FPGA} debugging},
  author={Attia, Sameh and Betz, Vaughn},
  booktitle={Proceedings of the ACM/SIGDA International Symposium on Field-Programmable Gate Arrays},
  pages={175--185},
  year={2020}
}

@inproceedings{sm_fpt,
author = {Attia, Sameh and Betz, Vaughn},
year = {2020},
month = {12},
pages = {206-214},
booktitle={2020 International Conference on Field-Programmable Technology (ICFPT)}, 
title = {State{R}eveal: {E}nabling {C}heckpointing of {FPGA} {D}esigns with {B}uried {S}tate},
doi = {10.1109/ICFPT51103.2020.00036}
}

@INPROCEEDINGS{RFUZZ,
  author={Laeufer, Kevin and Koenig, Jack and Kim, Donggyu and Bachrach, Jonathan and Sen, Koushik},
  booktitle={2018 IEEE/ACM International Conference on Computer-Aided Design (ICCAD)}, 
  title={RFUZZ: Coverage-Directed Fuzz Testing of RTL on FPGAs}, 
  year={2018},
  volume={},
  number={},
  pages={1-8},
  doi={10.1145/3240765.3240842}}

@INPROCEEDINGS{DifuzzRTL,
  author={Hur, Jaewon and Song, Suhwan and Kwon, Dongup and Baek, Eunjin and Kim, Jangwoo and Lee, Byoungyoung},
  booktitle={2021 IEEE Symposium on Security and Privacy (SP)}, 
  title={DifuzzRTL: Differential Fuzz Testing to Find CPU Bugs}, 
  year={2021},
  volume={},
  number={},
  pages={1286-1303},
  doi={10.1109/SP40001.2021.00103}}

@INPROCEEDINGS{ProcessorFuzz,
  author={Canakci, Sadullah and Rajapaksha, Chathura and Delshadtehrani, Leila and Nataraja, Anoop and Taylor, Michael Bedford and Egele, Manuel and Joshi, Ajay},
  booktitle={2023 IEEE International Symposium on Hardware Oriented Security and Trust (HOST)}, 
  title={ProcessorFuzz: Processor Fuzzing with Control and Status Registers Guidance}, 
  year={2023},
  volume={},
  number={},
  pages={1-12},
  doi={10.1109/HOST55118.2023.10133714}}

@inproceedings {Cascade,
author = {Flavien Solt and Katharina Ceesay-Seitz and Kaveh Razavi},
title = {Cascade: {CPU} Fuzzing via Intricate Program Generation},
booktitle = {33rd USENIX Security Symposium (USENIX Security 24)},
year = {2024},
isbn = {978-1-939133-44-1},
address = {Philadelphia, PA},
pages = {5341--5358},
url = {https://www.usenix.org/conference/usenixsecurity24/presentation/solt},
publisher = {USENIX Association},
month = aug
}

@misc{afl-fuzz,
  author       = {Micha{\l} Zalewski},
  title        = {{American Fuzzy Lop - Whitepaper}},
  year         = {2016},
  howpublished = {\url{https://lcamtuf.coredump.cx/afl/technical_details.txt}},
  note         = {Accessed: 2025-04-12}
}

@techreport{rocket,
  author      = {Krste Asanovic and Rimas Avizienis and Jonathan Bachrach and et al.},
  title       = {The Rocket Chip Generator},
  institution = {University of California, Berkeley},
  number      = {UCB/EECS-2016-17},
  year        = {2016},
  address     = {Berkeley, CA},
  type        = {Technical Report},
  url         = {https://www2.eecs.berkeley.edu/Pubs/TechRpts/2016/EECS-2016-17.html},
  note        = {Accessed: 2025-04-12}
}

@INPROCEEDINGS{Hyperfuzzing,
  author={Muduli, Sujit Kumar and Takhar, Gourav and Subramanyan, Pramod},
  booktitle={2020 IEEE/ACM International Conference On Computer Aided Design (ICCAD)}, 
  title={HyperFuzzing for SoC Security Validation}, 
  year={2020},
  volume={},
  number={},
  pages={1-9},
  doi={}}

@inproceedings{Effective_processor_verification,
author = {Kabylkas, Nursultan and Thorn, Tommy and Srinath, Shreesha and Xekalakis, Polychronis and Renau, Jose},
title = {Effective Processor Verification with Logic Fuzzer Enhanced Co-simulation},
year = {2021},
isbn = {9781450385572},
publisher = {Association for Computing Machinery},
address = {New York, NY, USA},
url = {https://doi.org/10.1145/3466752.3480092},
doi = {10.1145/3466752.3480092},
booktitle = {MICRO-54: 54th Annual IEEE/ACM International Symposium on Microarchitecture},
pages = {667–678},
numpages = {12},
location = {Virtual Event, Greece},
series = {MICRO '21}
}

@inproceedings {TheHuzz,
	author = {Rahul Kande and Addison Crump and Garrett Persyn and Patrick Jauernig and Ahmad-Reza Sadeghi and Aakash Tyagi and Jeyavijayan Rajendran},
	title = {{TheHuzz}: Instruction Fuzzing of Processors Using {Golden-Reference} Models for Finding {Software-Exploitable} Vulnerabilities},
	booktitle = {31st USENIX Security Symposium (USENIX Security 22)},
	year = {2022},
	isbn = {978-1-939133-31-1},
	address = {Boston, MA},
	pages = {3219--3236},
	url = {https://www.usenix.org/conference/usenixsecurity22/presentation/kande},
	publisher = {USENIX Association},
	month = aug
}

@misc{zhang2022optopenpretrainedtransformer,
      title={OPT: Open Pre-trained Transformer Language Models}, 
      author={Susan Zhang and Stephen Roller and Naman Goyal and Mikel Artetxe and Moya Chen and Shuohui Chen and Christopher Dewan and Mona Diab and Xian Li and Xi Victoria Lin and Todor Mihaylov and Myle Ott and Sam Shleifer and Kurt Shuster and Daniel Simig and Punit Singh Koura and Anjali Sridhar and Tianlu Wang and Luke Zettlemoyer},
      year={2022},
      eprint={2205.01068},
      archivePrefix={arXiv},
      primaryClass={cs.CL},
      url={https://arxiv.org/abs/2205.01068}, 
}

@inbook{genhuzz,
author = {Wu, Lichao and Rostami, Mohamadreza and Li, Huimin and Rajendran, Jeyavijayan and Sadeghi, Ahmad-Reza},
title = {GenHuzz: an efficient generative hardware fuzzer},
year = {2025},
isbn = {978-1-939133-52-6},
publisher = {USENIX Association},
address = {USA},
booktitle = {Proceedings of the 34th USENIX Conference on Security Symposium},
articleno = {93},
numpages = {19}
}

@INPROCEEDINGS{chatfuzz,
  author={Rostami, Mohamadreza and Chilese, Marco and Zeitouni, Shaza and Kande, Rahul and Rajendran, Jeyavijayan and Sadeghi, Ahmad-Reza},
  booktitle={2024 Design, Automation \& Test in Europe Conference \& Exhibition (DATE)}, 
  title={Beyond Random Inputs: A Novel ML-Based Hardware Fuzzing}, 
  year={2024},
  volume={},
  number={},
  pages={1-6},
  doi={10.23919/DATE58400.2024.10546625}}


\end{document}